\documentstyle[sprocl]{article}

\def\be{\begin{equation}}
\def\ee{\end{equation}}
\def\bea{\begin{eqnarray}}
\def\eea{\end{eqnarray}}
\newcommand{\beq}{\begin{equation}}
\newcommand{\eeq}{\end{equation}}
\newcommand{\beqs}{\begin{eqnarray}}
\newcommand{\eeqs}{\end{eqnarray}}

\begin{document}

\title{On the Large-$N_c$ Limit and Electroweak Interactions:
Some Properties of the $N_c$-Extended Standard Model\footnote{Talk given at
the Workshop on the Phenomenology of Large-$N_c$ QCD, Arizona State Univ.,
Jan. 2002; to be published in the proceedings}} 

\author{R. Shrock}

\address{C. N. Yang Institute for Theoretical Physics \\ State University of
New York \\ Stony Brook, NY 11794 \\E-mail: robert.shrock@sunysb.edu}

\maketitle\abstracts{We discuss properties of QCD with variable and large
$N_c$, taking into account electroweak interactions; i.e., we analyze the
generalization of the standard model based on the gauge group $G={\rm SU}(N_c)
\times {\rm SU(2)}_L \times {\rm U(1)}_Y$.  General classes of solutions to
anomaly constraints are given, and it is shown that these allow a $T_3 = \pm
1/2$ quark or $T_3=-1/2$ lepton rather than the neutrino as an electrically
neutral fermion.  The issue of grand unification is addressed, and it is shown
that $G$ cannot be embedded in the usual way in a simple SO($N$) or SU($N$)
gauge group unless $N_c=3$.  The ratio of strengths of QCD and electroweak
interactions is discussed.}

\section{Introduction}

The large-$N$ limit has long been of use in both statistical mechanics and
field theory.  Early examples of its application include the exact solutions of
an O($N$)-invariant spin model by Stanley \cite{stanley}, 2D U($N_c$) QCD by 't
Hooft \cite{thooft}, 2D models with four-fermion interactions \cite{gn} and
O($N$)-invariant scalar interactions~\cite{cjp}, and the O($N$)-invariant
nonlinear sigma model in $d=2+\epsilon$ dimensions~\cite{saclay,bls}.  The
large-$N_c$ expansion has been of great value as an analytic approach to the
nonperturbative properties of 4D QCD \cite{witten,sksb}; work continues on
properties of baryons, on finite-temperature and density behavior, and other
topics (reviews include \cite{jenkins,lebed}).

In applications of the large-$N_c$ expansion to QCD, a common practice
has been to turn off electroweak interactions and analyze the QCD sector by
itself.  However, it is of considerable interest to take into account the
electroweak interactions in the context of the large-$N_c$ limit and, more
generally, to investigate the properties of the full $N_c$-extended standard
model (SM) defined by the gauge group
\beq
G = {\rm SU}(N_c) \times {\rm SU}(2) \times {\rm U}(1)_Y 
\label{g}
\eeq
with fermions transforming as
\beq
Q_{iL} = \left (\begin{array}{c}
                  u_i \\
                  d_i \end{array} \right )_L \ : \ (N_c,2,Y_{Q_L})
\label{ql}
\eeq
\beq
u_{iR} \ : \ (N_c,1,Y_{u_R}) \ , \quad d_{iR} \ : \ (N_c,1,Y_{d_R})
\label{udr}
\eeq
\beq
{\cal L}_{iL} = \left (\begin{array}{c}
                  \nu_i \\
                   e_i \end{array} \right )_L \ : \ (1,2,Y_{{\cal L}_L})
\label{ll}
\eeq
\beq
\nu_{jR} \ : \ (1,1,Y_{\nu_R}) \ , \quad e_{iR} \ : \ (1,1,Y_{e_R})
\label{nuer}
\eeq
where $i$ denotes generation, $i=1,..N_{gen.}$, with $u_1 = u$, $u_2=c$,
$u_3=t$, $d_1=d$, $d_2=s$, $d_3=b$, etc.  Here $N_c$ is not necessarily large.
For generality, we shall consider arbitrary $N_{gen.}$.  We have analyzed this
theory \cite{nc} and report additional results here. 

In this $N_c$-extended standard model the usual relations $Q=T_3+Y/2$,
$Y_{Q_L}=q_u+q_d$, $q_u=q_d+1$, $q_\nu=q_e+1$, $Y_{f_R}=2q_{f_R}$ and the
vectorial property of electric charge, $q_{f_L}=q_{f_R} \equiv q_f$ continue to
hold. Hence $Y_{u_R}=Y_{Q_L}+1$, $Y_{d_R}=Y_{Q_L}-1$, $Y_{\nu_R}=Y_{{\cal
L}_L}+1$, and $Y_{e_R}=Y_{{\cal L}_L}-1$.  Before imposing anomaly cancellation
conditions, there are thus two independent electric charges among the fermions;
we take these to be $q_d$ and $q_e$.

In the general solution to the anomaly cancellation conditions for $N_c \neq
3$, the electric charges of all of the fermions will differ from their $N_c=3$
values.  In particular, since $q_{\nu} \ne 0$ in general, the theory should
include electroweak-singlet neutrinos $\nu_{jR}$ in order to form Dirac mass
terms $[\bar \nu_L \nu_R]$; for $q_\nu \ne 0$, Majorana mass terms would
violate electric charge conservation.  This contrasts with the situation in the
standard model where, because $\nu_{jR}$'s are singlets under $G_{SM}$, they
can be, and are, excluded from the fermion content. Hence, in the context of
the standard model as a renormalizable QFT with only operators of dimension
$\le 4$, there are no neutrino mass terms.  Of course, in the context of a
grand unified theory (GUT) such as SO(10), one does include $\nu_{jR}$'s with
$j=1,...N_{gen.}$.  The current evidence for neutrino masses and lepton mixing
also motivates the inclusion of heavy $\nu_{jR}$' to drive a seesaw mechanism
and provide an explanation for small neutrino masses.

In passing, we note that a more restricted $N_c$-extension of the standard
model avoids $\nu_{jR}$'s \cite{chowyan}.  There are then no gauge-invariant
renormalizable terms in the Lagrangian that can give neutrino masses, so in the
SM as a renormalizable QFT, the neutrinos are massless, a prediction disfavored
by current experimental indications.  In this restricted extension, one keeps
$q_\nu=0$ to avoid massless unconfined fermions. Here we shall consider the
full generalization with $\nu_{jR}$'s.

The electroweak part of the gauge group $G$ is still $G_{EW} = {\rm SU}(2) 
\times  {\rm U}(1)_Y$, and the issues of electroweak symmety breaking (EWSB)
are similar in the $N_c$-extended SM and the SM itself.  

\section{Anomaly Cancellation Conditions}

   The absence of anomalies is a necessary property of an acceptable quantum
field theory.  For the usual $d=4$ dimensional spacetime considered here, there
are three types of possible anomalies: (i) triangle anomalies in gauged
currents which, if present, would spoil current conservation and
renormalizability; (ii) the global SU(2) Witten anomaly resulting from the
nontrivial homotopy group $\pi_4(SU(2))=Z_2$ which, if present, would
render the path integral ill-defined; and (iii) mixed gauge-gravitational
anomalies (if one includes gravity).  We discuss here the constraints due to
the absence of these anomalies.  For anomalies (i), (iii), we suppress
generation index since the cancellation occurs separately for each generation. 

\subsection{Anomalies in Gauged Currents} 

The SU($N_c)^3$ anomaly vanishes automatically because of the vectorial nature
of the gluon-fermion couplings and the SU($N_c)^2$ U(1)$_Y$ anomalies vanish
automatically because of the vectorial nature of the electromagnetic couplings.
The condition for the vanishing of the SU(2)$^2$ U(1)$_Y$ anomaly is
\beq
N_c Y_{Q_L} + Y_{{\cal L}_L}=0 \ , \quad i.e. \quad 
N_c(2q_d+1)+(2q_e+1)=0
\label{yeq}
\eeq
The condition for the vanishing of the $U(1)_Y^3$ anomaly is
\beq
N_c(2Y_{Q_L}^3 - Y_{u_R}^3 - Y_{d_R}^3) +
(2Y_{{\cal L}_L}^3 - Y_{\nu_R}^3 - Y_{e_R}^3) = 0
\label{u1cubed}
\eeq
This yields the same condition as for the vanishing of the SU(2)$^2$ U(1)$_Y$
anomaly. Solving the equation for the vanishing of this anomaly gives, say for
$q_d$ in terms of $q_e$,
\beq
q_d = q_u -1 = -\frac{1}{2}\biggl ( 1 + \frac{Y_{{\cal L}_L}}{N_c} \biggr ) =
-\frac{1}{2}\biggl ( 1 + \frac{1}{N_c}(2q_e+1) \biggr )
\label{qdsol}
\eeq
or equivalently, taking $q_d$ as the independent variable,
\beq
q_e = q_\nu -1 = -\frac{1}{2}\biggl ( 1 + N_c Y_{{\cal L}_L} \biggr ) = 
-\frac{1}{2}\biggl ( 1 + N_c(2q_d+1) \biggr )
\label{qesol}
\eeq

\subsection{Global SU(2) Anomaly}

The constraint from the Witten global SU(2) anomaly is that the number
$N_{db.}$ of SU(2) doublets, $N_{db.} = (1+N_c)N_{gen.}$ is even.  We consider,
{\it a priori} the possibility of even and odd $N_{gen.}$.  If $N_{gen.}$ is
odd, the absence of the global SU(2) anomaly implies that $N_c$ is odd.  In
this case, there is a connection between $N_{gen.}$ and $N_c$.  In contrast, if
$N_{gen.}$ is even, there is no constraint on $N_c$.  Of course, if $N_c$ were
even, baryons would be bosons, and the properties of the world would be quite
different than for odd $N_c$.

\subsection{Mixed Gauge-Gravitational Anomalies}

The absence of mixed gauge-gravitational anomalies does not add any further
constraints; the mixed gauge-gravitational anomaly involving two graviton
vertices and an $SU(N_c)$ or $SU(2)$ gauge vertex vanishes identically since
$Tr(T_a)=0$ where $T_a$ is the generator of a nonabelian group.  The anomaly
involving a $U(1)_Y$ vertex is proportional to
 \beq
N_c(2Y_{Q_L}-Y_{u_R}-Y_{d_R})+(2Y_{{\cal L}_L}-Y_{\nu_R}-Y_{e_R}) = 0
\label{ggu1}
\eeq
where the expression vanishes because of the vectorial nature of the
electromagnetic coupling.  Indeed, the two separate terms in parentheses each
vanish individually: $2Y_{Q_L}-Y_{u_R}-Y_{d_R}=0$ and $2Y_{{\cal
L}_L}-Y_{\nu_R}-Y_{e_R}=0$, so that this anomaly does not connect quark and
lepton sectors. Hence, for a given $N_c$, there is a one-parameter family of
solutions for the fermion charges.  The values of these charges are real, but
not, in general, rational numbers, so that electric and hypercharge are not
quantized, although if one is rational, then all are, as is clear from
(\ref{yeq}).  In the SM extension with no $\nu_{jR}$'s and $q_\nu$ fixed at
zero, one does get charge quantization: $q_e=-1$, $q_d = q_u-1 = (1/2) (-1 +
N^{-1}_c)$.

\section{Issue of Grand Unification} 

An important question is whether one can embed the $N_c$-extended standard
model with gauge group $G$ in a grand unified theory (GUT) based on a simple
group $G_{GUT}$.  We recall that grand unification is appealing since it 
unifies quarks and leptons, predicts the relative sizes of gauge couplings 
in the SM factor groups, and quantizes electric charge, since $Q$ is a 
generator of $G_{GUT}$. Clearly,
\beq
{\rm rank}(G_{GUT}) \ge {\rm rank}(G) = N_c + 1
\label{rankgut}
\eeq
Following the standard procedure for constructing a GUT, one puts the fermions
in complex representations to avoid bare mass terms that would produce masses
of order $M_{GUT}$ for all fermions and requires no anomalies of types
(i)-(iii) in the theory.  The absence of mixed gauge-gravitational anomalies is
automatic, since there are no U(1) factor groups in $G_{GUT}$. Ideally, one
places all fermions of a given generation in a single irreducible
representation of $G_{GUT}$, although we shall also consider weakening this
condition. 

In order to place all fermions of each generation into one representation, a
necessary condition is that Tr$(Y) = \sum_f Y_f =0$ (for each generation) since
the hypercharge $Y$ is a generator of $G_{GUT}$.  This condition is satisfied:
\beq
{\rm Tr}(Y)= 2(N_c Y_{Q_L}+Y_{{\cal L}_L})+
N_c(Y_{u_R}+Y_{d_R})+Y_{e_R}+Y_{\nu_R} = 0
\label{try}
\eeq
Since the exceptional groups have bounded ranks, they cannot satisfy the rank
condition above for arbitrary $N_c$ and are excluded as candidates for
$G_{GUT}$ for general $N_c$.  To guarantee the absence of anomalies in gauged
currents, one uses a ``safe'' group, for which the generators satisfy $A_{abc}
= {\rm Tr}(\{T_a,T_b\},T_c)=0 \ \forall \ a,b,c$.  Recall that SU($N$) is not
safe for $N \ge 3$, SO($N$) has only real representations for odd $N$ and for
$N=0$ mod 4, while SO($N$) has complex representations and is safe for $N=2$
mod 4, except for $N=6$ (SO(6) $\simeq$ SU(4)). This leads to the choice
$G_{GUT}=$ SO($4k+2$) as the GUT group in which to embed $G$.  This is the
natural generalization of the SO(10) GUT \cite{so10} in which the SM gauge
group $G_{SM}={\rm SU}(3)_c \times {\rm SU}(2)_L \times U(1)_Y$ is embedded.
There are then no triangle anomalies in gauged currents and also no global
anomaly, since $\pi_4({\rm SO}(N))=\emptyset$ for $N \ge 6$.  Now,
rank(SO($2n))=n$, so rank(SO($4k+2))=2k+1$.  Substituting into the inequality
rank($G_{GUT}) \ge N_c+1$ yields $2k \ge N_c$.  Consider first the case of odd
$N_{gen.}$.  For this case, $N_c$ must be odd for the absence of the global
SU(2) anomaly in $G$, so the above inequality becomes $2k \ge N_c+1$, and hence
the minimal-rank GUT group is ${\rm SO}(4k+2)={\rm SO}(2N_c+4)$ with rank
$N_c+2$.  Since $N_c+2$ is odd, $2N_c+4 = 2$ mod 4, and ${\rm SO}(2N_c+4)$ has
a complex spinor representation of dimension $2^{N_c+1}$.  For each generation,
there are $N_f = 4(N_c+1)$ Weyl fermions.  The condition that these fit into a
spinor is then
\beq
2^{N_c+1} = 4(N_c+1)
\label{ffit}
\eeq
{\it But this has a solution only for $N_c=3$.} 

In the hypothetical case that $N_{gen.}$ is even, then the global SU(2) anomaly
condition allows $N_c$ to be either even or odd.  The case of odd $N_c$ has
been covered.  For even $N_c$, the minimum-rank $G_{GUT}$ is ${\rm
SO}(2N_c+2)$.  Since $N_c$ is even, $2N_c+2 = 2$ mod 4, and ${\rm SO}(2N_c+2)$
has a complex spinor of dimension $2^{N_c+1}$, so that one is again led to the
same condition (\ref{ffit}) and conclusion.  This is an important result
\cite{nc}, since it shows that one can grand-unify the $N_c$-extended standard
model as discussed above only for the single case $N_c=3$.

One could also attempt a less complete type of grand unification, in which one
assigns fermions of a given generation to more than one representation of
$G_{GUT}$. Since SU($N_{GUT}$) is not a safe group for $N_{GUT} \ge 3$, one
must arrange for the triangle anomaly in gauged currents to cancel between
representations.  Since fermions are not placed in a single representation, the
conditions ${\rm Tr}(Y)=0$ and ${\rm Tr}(T_3)=0$ or ${\rm Tr}(Q)=0$ are also
not automatically satisfied for each representation, as they would be if all
the fermions of a given generation are placed in a single representation of
$G_{GUT}$.  

As a generalization of the Georgi-Glashow SU(5) GUT \cite{su5}, we fix
$q_\nu=0$ so that one can exclude $n_{jR}$'s, and try to fit the remaining
$4(N_c+1)-1=4N_c+3$ Weyl fermions of each generation in some conjugate
fundamental representations $(\psi_{k,L}^c)_\alpha$ and an antisymmetric
second-rank tensor rep. $\psi_{L}^{[\alpha \beta]}$ of SU($N_{GUT}$), with
$\alpha,\beta=1,...N_{GUT}$.  Since the contribution to the triangle anomaly in
gauged currents from $\psi_{L}^{[\alpha \beta]}$ is $(N_{GUT}-4)$ times that
from the fundamental representation, $(\psi_L^{\alpha})$, we use $N_{GUT}-4$
copies of $(\psi_{k,L}^c)_\alpha$, with $k=1,2,..,N_{GUT}-4$.  The condition
that the fermions of each generation fit in these representations is then
\beq
4N_c+3 = (N_{GUT}-4)N_{GUT} + \frac{N_{GUT}(N_{GUT}-1)}{2} = 
\frac{3N_{GUT}(N_{GUT}-3)}{2}
\label{suncond}
\eeq
The solution is 
\beq
N_{GUT} = \frac{1}{2} \biggl [ 3 + \frac{1}{3}\sqrt{3(51+32N_c)} \ \ \biggl ]
\label{ngutsol}
\eeq
for integer $N_{GUT}$.  As before, one also requires
\beq
{\rm rank}(G_{GUT}) \ge {\rm rank}(G) = N_c+1
\label{rankrel}
\eeq
i.e.,
\beq 
\Delta _{rank} = {\rm rank}(G_{GUT})- {\rm rank}(G) = N_{GUT} - N_c -2 \ge 0
\label{deltarank}
 \eeq
Again, the only solution is $N_c=3$, $N_{GUT}=5$, i.e., SU(5), with
$\Delta_{rank}=0$ \cite{su5}.  The next two solutions with integer values of
$N_{GUT}$ are
\beq
(N_c,N_{GUT},\Delta_{rank}) = (6,6,-2) \ , \ (48,13,-37)
\label{ncngsols}
\eeq
and both are excluded by the negative $\Delta_{rank}$, which becomes
increasingly negative for larger $N_c$.  Thus, one reaches a negative
conclusion even before addressing whether the above trace conditions could be
satisfied.  So even if one attempts this less ambitious type of grand
unification of the group (\ref{g}) with general $N_c$, it is only possible for
the special case $N_c=3$.  Of course these results do not reduce the usefulness
of the large-$N_c$ expansion in pure QCD.  However, they do show how special
the value $N_c=3$ is from the point of view of grand unification.

\section{Classification of Solutions for Fermion Charges}

In the general $N_c$-extended standard model, there are several generic and
special classes of solutions for the fermion charges.  The classes, denoted
$Cn_q$, for the quark charges are given in the table.  Here classes $C1_1$,
$C2_q$, and $C3_q$ are generic, while $C2_{q,sym}$ is a symmetric subcase of
$C2$ and $C4_q$ and $C5_q$ are other special solutions.  The charges $q_u$ and
$q_d$ are monotonically increasing functions of $N_c$ if $q_e < -1/2$ and
monotonically increasing functions of $N_c$ if $q_e > -1/2$.  In the borderline
case $q_e=-q_\nu=-1/2$ (whence $Y_{{\cal L}_L}=0$), $q_u$ and $q_d$ are
independent of $N_c$ and have the values in $C2_{q,sym}$, $q_d=-q_u=-1/2$
(whence $Y_{Q_L}=0$), so that the anomalies of type (i) cancel separately in
the quark and lepton sectors.

\begin{table}[t]
\caption{Classes of solutions for quark charges} 
\vspace{0.2cm}
\begin{center}
\begin{tabular}{|c|c|c|c|c|} \hline \hline & & & & \\
case & $q_d$ & $(q_u,q_d)$ & $Y_{Q_L}$ & $Y_{{\cal L}_L}$ \\
& & & & \\
\hline \hline
$C1_q$  & $> 0$          & $(+,+)$ & $> 1$ & $< -N_c$ \\ \hline
$C2_q$  & $-1 < q_d < 0$ & $(+,-)$ & $-1 < Y_{Q_L} < 1$ & $-N_c < Y_{{\cal
                                                    L}_L} < N_c$ \\ \hline
$C2_{q,sym}$ & $-1/2$ & $(1/2,-1/2)$  & 0 & 0 \\ \hline
$C3_q$  & $< -1$     & $(-,-)$ & $< -1$ & $> N_c$ \\ \hline
$C4_q$ & 0    & (1,0)    & 1 & $-N_c$ \\ \hline
$C5_q$ & $-1$ & $(0,-1)$ & $-1$ & $N_c$ \\ \hline
\hline
\end{tabular}
\end{center}
\label{table1}
\end{table}

One can also work out analogous classes of solutions $C1_\ell$ to $C5_\ell$ for
lepton charges satisfying the anomaly constraints \cite{nc}.  These results
show that the solutions to the anomaly cancellation conditions allow a $T_3 =
\pm 1/2$ quark or $T_3=-1/2$ lepton rather than the neutrino as an electrically
neutral fermion.

\section{Relative Strengths of Color and Electroweak Interactions} 

As the above solutions show, if $Y_{Q_L} \ne 0$, then the lepton charges $q_e$
and $q_\nu$ will diverge as $N_c \to \infty$: $q_e, q_\nu \sim -Y_{Q_L}N_c/2$.
A necessary condition for the lepton charges to remain finite as $N_c \to
\infty$ is that $\lim_{N_c \to \infty} q_d = -1/2$, i.e., $\lim_{N_c \to
\infty}Y_{Q_L}=0$.  In contrast, for any fixed finite value of $q_e$, the quark
charges $q_u$ and $q_d$ have finite limits as $N_c \to \infty$: $\lim_{N_c \to
\infty} q_d = -1/2$, $\lim_{N_c \to \infty} q_u = 1/2$.

Even if all of the fermion charges remain finite as $N_c \to \infty$,
electroweak effects that involve quarks in loops can still lead to divergent
behavior in this limit. Some examples are provided by quark loop corrections to
the $W$, and $Z$ propagators, which are $\propto N_c$.  To keep these finite in
the $N_c \to \infty$ limit, it is sufficient to require that the SU(2) and
U(1)$_Y$ gauge couplings $g$ and $g'=(3/5)^{1/2}g_1$ satisfy $g^2N_c = c_2$ and
$g^{\prime 2} N_c = c_Y$ analogous to the condition on the color SU(3)$_c$
coupling $(g_s)^2 N_c = c_3$.  From the relation $e=gg'/\sqrt{g^2+g'^2}$, it
follows that $e^2 N_c = {\rm const.}$, which keeps the quark-loop contributions
to the photon propagator finite.  The electromagnetic corrections to the baryon
mass arising from one-photon exchange between quark lines then go like $(\delta
m_B)_{em} \propto e^2 N_c^2 \propto N_c$, which is the same as the behavior
$m_B \propto N_c$ in the QCD sector.

An example of a ratio of electroweak cross sections that involves $N_c$ is
$\sigma(e^+e^- \to \mu^+\mu^-)/\sigma(e^+e^- \to {\rm hadrons})$ at a
center-of-mass energy $\sqrt{s} >>$ hadron masses (i.e. quite high, given that
$m_B \sim N_c$) and away from the positions of narrow meson states, so that
scaling can work:
\beq
\frac{\sigma(e^+e^- \to \mu^+\mu^-)}{\sigma(e^+e^- \to {\rm hadrons})} \sim
\frac{1}{N_c}
\label{rratio}
\eeq

A different issue concerning the strength of electroweak interactions arises
for fixed but arbitrary $N_c$: even if the gauge couplings $g, g'$ are small,
the one-parameter solutions for the fermion charges allow arbitrarily large
values of these charges and similarly for hypercharges. Hence, the actual
strengths of the U(1)$_Y$ and electromagnetic interactions, $|g'Y_f |$ and $|e
q_f|$ can be arbitrarily large.  One can avoid this problem by setting some
finite (not necessarily zero) input value for $q_\nu$.

Part of the evidence for grand unification in our actual world is that, even
before calculating precision evolution of the gauge couplings, one has the
requisite ordering; at a given scale $\mu$, of the $3!=6$ possible relative
orderings in size of the SU(3), SU(2), and U(1) couplings $g_s^2$, $g^2$, and
$g_1^2$, one has $g_s^2 > g^2 > g_1^2$.  For $N_c \ne 3$, since we do not have
a usual GUT, different orderings might, {\it a priori}, occur, and could
produce quite different theories.  For example, for fixed (moderate) $N_c$,
consider the possibility that $g^2 >> g_s^2, \ g_1^2$.  In this case, for
sufficiently large $g^2/(4 \pi) \sim O(1)$, the SU(2) gauge interaction could
produce the following SU(2)-invariant condensates breaking color- and/or 
charge as well as baryon and/or lepton number: 
\beq
\langle \epsilon_{ab} Q^{\alpha a T}_{iL} C Q^{\beta b}_{jL} \rangle = 
\langle u^{\alpha T}_{iL} C d^{\beta}_{jL} - 
d^{\alpha T}_{iL} C u^{\beta}_{jL} \rangle
\label{con1}
\eeq
\beq
\langle \epsilon_{ab} Q^{\alpha a T}_{iL} C {\cal L}^b_{jL} \rangle =
\langle u^{\alpha T}_{iL} C e_{jL} - 
d^{\alpha T}_{iL} C \nu_{jL} \rangle
\label{con2}
\eeq
\beq
\langle \epsilon_{ab} {\cal L}^{a T}_{iL} C {\cal L}^b_{jL} \rangle =
\langle \nu^T_{iL} C e_{jL} - e_{iL} C \nu_{jL} \rangle
\label{con3}
\eeq
where $a$, $\alpha$, and $(i,j)$ are SU(2), SU($N_c$), and generation indices,
respectively.  This shows that even though QCD and electromagnetism are
vectorial gauge symmetries, they could be broken because of the SU(2)$_L$ 
interaction.   This is prevented if $g^2$ is sufficiently small.  

\section{Some Properties of Bound States}

First, consider baryons composed of $r$ up-type and $N_c-r$ down-type quarks;
these have electric charge $q[B(r,N_c-r)]=r+N_c q_d$.  The charge difference
$q[B(r,N_c-r)]- q[B(N_c-r,r)]=2r-N_c$. Now assume that $N_c$ is odd and that
$m_u, m_d << \Lambda_{QCD}$, as in the physical world. A strong--isospin mirror
pair which constitutes a kind of generalization of the proton and neutron is
\beq
{\cal P} = B \biggl ( \frac{N_c+1}{2}, \frac{N_c-1}{2} \biggr ) \ , \quad 
{\cal N} =B \biggl ( \frac{N_c-1}{2}, \frac{N_c+1}{2} \biggr )
\label{pn}
\eeq
These baryons have charges $q_{\cal P} =q_{\cal N}+1$ satisfying
$q_{\cal P} = -q_e$ and $q_{\cal N}=-q_\nu$. 

One can also consider atoms.  For all cases except $C4_\ell$ ($q_e=0$), there
exists a neutral Coulomb bound state of the generalized proton ${\cal P}$ and
electron, $({\cal P}e)$, which is the $N_c$-extended generalization of the
hydrogen atom.  For all cases except $C5_\ell$ ($q_\nu=0$), there also exists a
second neutral Coulomb bound state, which has no analogue in the usual $N_c=3$
standard model, namely, $({\cal N}\nu_{1})$, where $\nu_{1}$ denotes the
lightest neutrino mass eigenstate.  For our discussion of the $({\cal P}e)$ and
$({\cal N}\nu)$ atoms, we assume, respectively, that $q_e \ne 0$ and $q_\nu \ne
0$, so that these atoms exist, and we suppress the mass eigenstate index in
$({\cal N}\nu_1)$.

The $({\cal P}e)$ and $({\cal N}\nu)$ atoms are nonrelativistic bound states
iff $|q_e| \alpha_{em} << 1$ and $|q_\nu| \alpha_{em} << 1$ (equivalent
conditions).  Then the binding energy in the ground state of the $({\cal P}e)$
atom is $E_{({\cal P}e)} = -(q_e\alpha)^2 m_e/2$ and the Bohr radius is $a_0 =
1/(q_e^2\alpha m_e)$.  These formulas apply to $({\cal N}\nu)$ with the
replacements ${\cal P} \to {\cal N}$ and $e \to \nu$.  For large $N_c$, $e^2N_c
=$ const., and fixed $q_e$, the binding is very weak, $E_{({\cal P}e)}, \
E_{({\cal N}\nu)} \propto N_c^{-2}$ and $a_0 \propto N_c$. Since the size of
the nucleon $r_N \sim O(1)$ as $N_c \to \infty$, the electron and neutrino
clouds in these respective atoms extend over much larger distances than $r_N$.
Indeed, since $a_0$ diverges as $N_c \to \infty$, the notion of an individual
atom reasonably well separated from other atoms requires that interatomic
separations grow at least like $N_c$.  

For large $N_c$ the strong interactions of mesons become weak and these mesons
are long-lived.  Hence, one could also consider meson-lepton Coulomb bound
states.  Consider, for example, the $J=0$ $u \bar d$ meson $\pi^+$ (or the
$J=1$ $u \bar d$ meson $\rho^+$).  Possible bound states of
$\pi^+$ with leptons include (i) $(\pi^+ e)$, if $q_e < 0$ as in cases
$C2_\ell$ and $C3_\ell$; this would have charge $q=1+q_e$ and hence would be
neutral for the special case $C5_\ell$; (ii) $(\pi^+ \bar e)$ if $q_e > 0$ as
in case $C1_\ell$; this would have charge $q=1-q_e$; (iii) $(\pi^+ \nu)$ if
$q_\nu < 0$ (case $C3_\ell$) with charge $q=1+q_\nu$; and (iv) 
$(\pi^+ \bar\nu)$ if $q_\nu > 0$ (cases $C1_\ell$, $C2_\ell$) with charge
$1-q_\nu$, hence neutral for case $C4_\ell$. 

There might also be a stable purely leptonic Coulombic bound state with lepton
number $L=2$, $(e \nu)$. This requires that $q_e \ne 0$, $q_\nu \ne 0$, and
${\rm sgn}(q_e)=-{\rm sgn}(q_\nu)$, which happens in case $C2_\ell$.  Unlike
the neutral atoms $({\cal P}e)$ and $({\cal N}\nu)$, this possible leptonic
state would, in general, be charged:

In the usual world, molecules are stable because the Coulomb repulsion of the
nuclei is counterbalanced by the Coulombic attraction between nuclei and the
total set of electrons, yielding various bonds (ionic, covalent).  For large
$N_c$ this balancing could still occur, yielding molecules and normal
crystalline matter, but on an interatomic length scale $r_{i.a.} \sim N_c$.

\section*{Acknowledgments}

I would like to thank R. Lebed for the organization of this conference, which
was also aided by the Institute for Nuclear Theory, Univ. of Washington. 
The research of R. S. is partially supported by the NSF grant PHY-0098527.


\section*{References}
\begin{thebibliography}{99}

\bibitem{stanley}{H. E. Stanley, Phys. Rev. {\bf 176}, 718 (1968).}

\bibitem{thooft}{G. 't Hooft, Nucl. Phys. {\bf B72}, 461 (1974); {\it ibid},
{\bf B75}, 461 (1974).}

\bibitem{gn}{D. Gross and A. Neveu, Phys. Rev. {\bf D10}, 3235 (1974).} 

\bibitem{cjp}{S. Coleman, R. Jackiw, and H. D. Politzer, Phys. Rev. 
{\bf D10}, 2491 (1974).} 

\bibitem{saclay}{E. Br\'ezin and J. Zinn-Justin, Phys. Rev. {\bf B14}, 3110
(1976); E. Br\'ezin, J. Zinn-Justin, and J. Le Guillou, Phys. Rev. {\bf D14},
2615 (1976).}

\bibitem{bls}{W. Bardeen, B. W. Lee, and R. Shrock, Phys. Rev. {\bf D14}, 
985 (1976).}

\bibitem{witten}{E. Witten, Nucl. Phys. {\bf B160}, 57 (1979);
Nucl. Phys. {\bf B223}, 422, 433 (1983).}

\bibitem{sksb}{S. Coleman and E. Witten, Phys. Rev. Lett. {\bf 45}, 100
(1980).}

\bibitem{jenkins}{E. Jenkins, Ann. Rev. Nucl. Part. Sci. {\bf 48}, 81 (1998);
hep-ph/0111338.} 

\bibitem{lebed}{R. Lebed, nucl-th/9810080.}

\bibitem{nc}{R. Shrock, Phys. Rev. {\bf D53}, 6465 (1996).} 

\bibitem{chowyan}{C.-K. Chow and T.-M. Yan, Phys. Rev. {\bf D53}, 5105
(1996). See also A. Abbas, Phys. Lett. {\bf 238B}, 344 (1990).}

\bibitem{so10}{H. Georgi, in {\it Particles and Fields, 1974} (A.I.P., NY,
1975), p. 575; H. Fritzsch and P. Minkowski, Ann. Phys. {\bf 93}, 193 (1975).}

\bibitem{su5}{H. Georgi and S. Glashow, Phys. Rev. Lett. {\bf 32}, 438 (1974).}

\end{thebibliography}
\end{document}